\documentstyle[editedvolume]{crckapb} 

\begin{opening}
\title{GR 15 Session A5(ii) Algebraic computing}
\author{Kayll Lake} 
\institute{Department of Physics, Queen's University\\Kingston Ontario Canada K7L 3N6}
\end{opening}
\runningtitle{Computer Methods in GR: Algebraic computing}
\begin{document}
\section{Introduction}
Algebraic computing in relativity and gravitation dates back more than thirty years, but only relatively recently has hardware of sufficient power to tackle large scale calculations become commonplace. Whereas it is generally understood throughout the relativity community that there are a number of packages available, the diversity of problems that the available packages can help with is not so widely appreciated.  This session was devoted to computer algebra for relativity and gravitation from the point of view of developers. In this summary I expand this to include some background and outline what is available for users in the field. 

\section{Background}
If indeed "the beginning is the most important part of the work" \footnote{Plato, The Republic. Book II. 377B} let me start by emphasizing the fact that modern computer algebra systems are {\it easy} to use and can do rather more than calculate the components of the Riemann tensor! About fifteen years ago it would have been reasonable to list active users of computer algebra systems in the relativity community (e.g., \cite{Cohen84}). This is no longer feasible. And yet, for example, we find "Algebraic Computing in General Relativity" but one of six topics in "{\it Numerical} Relativity" within the newly established Living Reviews In Relativity. \footnote{The web address is www.livingreviews.org} To this day computer algebra is, I believe, an underused tool. It can certainly be argued that it is also under supported \cite{Petti94}, and perhaps even misunderstood. Three questions come to the fore: How did computer algebra develop, how is it distinct from more traditional uses of the computer, and what can it do for researchers in relativity and gravitation? 

 As regards the history,  useful  references are \cite{D'Inverno80} and  \cite{MacCallum96}. In computer algebra calculations, expressions tend to inflate before they simplify. \footnote{An example of this intermediate growth is provided by the Tomimatsu-Sato spacetimes. A discussion can be found at www.astro.queensu.ca/\char'176lake/demo.html and at www.scar.utoronto.ca/\char'176harper/redten/node60.html} This intermediate expression growth dictates garbage collection, a standard feature of LISP and C but not of FORTRAN. Moreover, since the final size of expressions are not known {\it a priori}, recursive algorithms, which are impossible in FORTRAN, are an essential feature of any computer algebra system. \footnote{ See e.g., \cite{MacCallum96}.} The consequence is that at  present the available algebraic engines  are LISP-like (e.g., REDUCE \footnote{ www.rrz.uni-koeln.de/REDUCE} and Macsyma \footnote{www.macsyma.com}) or C-like (e.g., MapleV \footnote{www.maplesoft.com} and Mathematica \footnote{www.wri.com}). Intermediate expressions must of course be stored and simplified. The key requirements of a computer algebra system are; in the software the ability to {\it simplify}, and from the hardware ample high speed {\it memory} (RAM). CPU speed is of some interest, but is considerably less important than it used to be. \footnote{A reasonable number to gauge hardware speed for computer algebra  is SPECint95 (www.specbench.org/osg/cpu95/results/cint95.html). At the time of writing, PCs are comparable to traditional workstations on this measure and the associated SPECint95 numbers have increased by about a factor of four in the last year and a half.} 

The environment that a computer algebra system presents to a user, familiar with numerical computing,  is strange territory, though perhaps not at first sight. The level of complexity of a problem submitted to the system is not always clear beforehand. Consider, for example, a calculation of the number of terms in the covariant Ricci tensor in n dimensions for a general form of the metric (that is, n(n+1)/2 functions of n variables) \cite{Fee88}. With n=4 the number of terms in {\it each} diagonal component of Ricci is 9,990 and in each off-diagonal component there are 13,280 terms. This once classic problem can now be handled on almost any PC with appropriate software. However, the number of terms grows rapidly with n. At n=5 the number of terms in each diagonal component of Ricci rises to 298,134, and to 410,973 in each off-diagonal component. The case n=5, without recourse to custom code,  requires a dedicated resource of about {\small 1/2}GB of RAM. \footnote{At the present time, within MapleV, this calculation also requires a 64-bit architecture, a feature one can expect of PCs in the near future.} A increase in difficulty from n=4 to n=5 is certainly anticipated, but the magnitude of this change is difficult to predict. Another feature essentially universal to all modern computer algebra systems is the need for effective use of the available simplification procedures. This requires some practice, though some general guidelines are usually available for  particular systems. \footnote{For example, www.astro.queensu.ca/\char'176lake/simp.html} For most situations these general rules will give adequate performance \footnote{Usually, it is the answer that is of interest and not the fact that the simplification strategy is optimal. When optimal simplification strategies are the prime concern the problem is more involved because of the large number of simplification procedures that are usually available and the size of the resultant parameter space to be explored.}, and reduce the calculations for even complex spacetimes to an essentially trivial exercise, though one that could not be attempted by hand. However, failure to follow the rules can make even simple calculations intractable. In practice, computer algebra calculations tend to be either trivial, or impossible. A computer algebra system is a tool, and the skill of the user is measured by the ability to turn the impossible into the trivial. In reality this skill is rather easy to obtain. 

What is available, and what features can a user hope to find? In an attempt to answer the first question I maintain a list.  \footnote{www.astro.queensu.ca/\char'176lake/course/algebra.html} The second question is harder to deal with. I outline below some thoughts on what one would like to see in a modern package: \footnote{A personal note of caution is in order here. The concept that you get what you pay for does not seem, at least to me, to extend to the software "industry".}
\begin{itemize}
\item    A transparent easy to use feature-rich interface;
\item    A full slate of simplification and manipulation routines;
\item    The ability to work in a variety of  formalisms with seamless integration; 
\item    A menu which contains all standard objects along with derivatives in any index configuration, operators, support for specialized symmetries, and Petrov and Segr\'e typing;
\item    The ability to define, in a transparent way, new objects taking into account their symmetries;
\item    The ability to work with multiple metrics simultaneously and determine junction conditions on timelike, spacelike and null boundaries; 
\item    The ability to develop customized packages (in any number of dimensions) that can be used in conjunction standard functions and objects;
\item    An on board database of known solutions of Einstein's equations with "equivalence problem" tools; 
\item    A robust tightly integrated abstract package;                                                              
\end{itemize}

Unfortunately, no single package is currently available which does all of the above.

\section{Presentations}
At the session the "equivalence problem" (the unique coordinate-free characterization of a space, see e.g., \cite{MacSkea94}) once again dominated the contributions. Jan \AA man \cite{Aman97} discussed a class of Riemann-Cartan G\"odel-type space-times from the point of view of equivalence problem techniques by using a suite of computer algebra programs called {\sc tclassi}. A coordinate-invariant description of the gravitational field for this class of space-times was presented and the possible groups $G_{r}$ of affine-isometric motions were discussed. Conditions for space-time (ST) local homogeneity were derived and the local equivalence of the ST homogeneous Riemann-Cartan G\"odel-type space-times was discussed. It was shown that they admit a five-dimensional group of affine-isometries and are characterized by three essential parameters $\,\ell, m^2, \omega\,$: identical triads ($\ell, m^2, \omega$) correspond to locally equivalent manifolds.The algebraic types of the irreducible parts of the curvature and torsion tensors are also presented. Denis Pollney \cite{Pollney97}  described a suite of programs to help in the calculation of an invariant classification of a space-time in MapleV.  The algorithm used is largely based on (though not equivalent to) the algorithm used in Jan{\AA}man's CLASSI extension of the SHEEP computer algebra system.  He  described the implementation of the classification algorithms in MapleV within the GRTensorII package where factorization, integration or graphical facilities are available. The basic algorithms employed have been documented and are to be placed  in the public domain. The method used for classification is that of  \cite{Karlhede81} which uses algebraic properties of the curvature spinors to align the basis vectors and restrict the SL(2,C) freedom of the tetrad as far as possible in order to reduce the number of independent components to be calculated. As he pointed out, an important consideration in implementations of this method is the choice of invariant canonical forms for the curvature spinor components and their derivatives.  A specific set of canonical forms for symmetric spinors was proposed, with a discussion of  alternatives, with particular attention to computability, practical use, and interpretation. To this end, he described new tools in GRTensorII
for the calculation and manipulation of symmetric spinors, including the calculation of their independent derivatives to any order as well as the ability to carry out general SL(2,C) rotations of their components. Ray d'Inverno \cite{d'Inverno97} presented an interactive database of exact solutions of Einstein's field equations available on the web. \footnote{Available at edradour.symbcomp.uerj.br, at www.maths.soton.ac.uk/\char'176rdi/database and at www.astro.queensu.ca/\char'176jimsk} The intention of the database is not simply to provide a list of exact solutions, but also to allow the user access to the invariant classification of these solutions. The database provides access to a unique characterization of the Riemann tensor and its covariant derivatives for over 200 metrics, which the user can view without performing calculations. The database allows three modes of search:  (i) by one or more discrete characteristics which the user selects by radio buttons, (ii) by keyword search, and (iii) by equation number as listed in \cite{Kramer80}. Future plans include an interactive interface to invariant classification routines to enable a user to generate invariant classifications "on-line".

In addition to discussions of  the "equivalence problem", Jim O'Connor \cite{O'Connor97} described use of  the package Dimsym \footnote{www.latrobe.edu.au/www/mathstats/Maths/Dimsym/} (a program primarily for the determination of symmetries of differential equations)  to seek collineations for the Kimura metrics. He demonstrate that when a Killing tensor is known for a metric one can seek an associated collineation by solving first order equations that give the Killing tensor in terms of the collineation rather than the second order determining equations for collineations. Anders H\"oglund \cite{Hoglund97}  described Tensign, a new program for indicial calculations of tensors. This program takes advantage of a cursor that you move around in the expression so that you can perform your operations just where you want them. One example of a calculation that has been done with Tensign is the wave equation for the Lanczos potential \cite{Edgar97}. Another example is the investigation of the existence of the Lanczos potential in dimensions larger than four \cite{Andersson97}. I described  some innovative work by Peter Musgrave \cite{Musgrave97} which is in effect the first (as far as I am aware) {\it virtual} algebraic calculator for the web. GRLite \footnote{www.astro.queensu.ca/\char'176lake/vac.html} is an experimental easy to use GUI to GRTensorII. The Java applet \footnote{In Pune I was able to demonstrate  GRLite (with the server in Canada)  on a modest computer (Intel 386@33MHz) by way of the Kretschmann scalar for the Kerr-Newman metric. When a connection was available, the demonstration took but a few seconds.} is a client which connects to a Java-based server. This server in turn starts a MapleV session and acts as an intermediary translating the button clicks and menu selections into GRTensorII commands. GRLite provides a prototype for a platform (hardware and algebraic engine) independent unified GUI to a variety of computer algebra programs on and off the Web. 

\section{Some Recent Examples}
Some recent examples, selected from preprints received in the last two weeks, demonstrate how computer algebra has been used very recently as a tool. Unruh \footnote{astro-ph/9802323, start at xxx.lanl.gov} has  reexamined cosmological long wavelength perturbations and presented an {\it exact} solution to the long wavelength perturbations for the scalar modes and for a scalar field theory with arbitrary potential.  The algebraic manipulations necessary in driving the equations and in reducing the Hamiltonian action were done with the help of a computer algebra system. The same author  has also recently examined radiation reaction fields for an accelerated dipole for scalar and electromagnetic radiation. \footnote{physics/9802047} The radiation reaction fields at the location of a point source for a massless scalar field, and the reaction fields at the location of a point dipole source for the electromagnetic field were derived using the Penrose integral. The extensive  calculations were aided in an essential way with a computer algebra system. Glass and Krisch \footnote{gr-qc/9803040} have extend the Vaidya radiating metric to include both a radiation field and a string fluid. Assuming diffusive transport for the string fluid, they find new analytic solutions of Einstein's field equations. These represent an extention of Xanthopoulos superposition. To end, I would like to point out some recent work by George Davies. \footnote{www.astro.queensu.ca/\char'176 lake/cbh.html} He has developed computer algebra methods for calculating black hole perturbations which allow the ``second order Zerilli function'' \footnote {Discussed elsewhere in these proceedings by Pullin. See gr-qc/9803005} to be calculated in but a couple of moments on a PC. Work is in progress to extend this analysis to the Kerr metric.


\begin{thebibliography}{}
\bibitem[\protect\citeauthoryear{Cohen {\it et al.}}{1984}]{Cohen84}
I. Cohen, I. Frick and J. E. Aman (1984) Algebraic Computing in General Relativity, in {\it General Relativity and Gravitation }, {\bf GR 10}, B. Bertotti, F. de Felice and A. Pascolini, editors, Reidel Publishing Company, Dordrecht,{pp.~139--162}
\bibitem[\protect\citeauthoryear{Petti}{1994}]{Petti94}
R. Petti (1994) Why Math Software Research and Development Counts {\it Computers in Physics}, {\bf Vol. 8, No. 6}, {pp.~623 }(www.macsyma.com/MIRoleOfMathSW.html)
\bibitem[\protect\citeauthoryear{D'Inverno}{1980}]{D'Inverno80}
R.A. D'Inverno (1980) A Review of Algebraic Computing in General Relativity, in {\it General Relativity and Gravitation, One Hundred Years After the Birth of Albert Einstein} A. Held, editor, Plenum Press, New York {pp.~491--537}\footnote{It is of historical interest to note that the fastest time for the Bondi test quoted in Table 3 (on dedicated mainframe computers of the era) is typically of the order of twenty times slower than a modest contemporary PC with  general purpose software.}
\bibitem[\protect\citeauthoryear{MacCallum}{1996}]{MacCallum96}
M.A.H. MacCallum (1996) Computer Algebra and Applications in Relativity and Gravity, in {\it  Recent Developments in Gravitation and Mathematical Physics: Proceedings of the First Mexican School on Gravitation and Mathematical Physics}, A. Macias, T. Matos, O. Obregon, and H. Quevedo, editors,  World Scientific, Singapore,
{pp.~3--41}
\bibitem[\protect\citeauthoryear{Fee {\it et al.}}{1988}]{Fee88}
G.J. Fee, R.G. McLenaghan and R. Pavelle (1988) GR12 Contributed Papers p296. For a discussion see www.astro.queensu.ca/\char'176lake/demo.html
\bibitem[\protect\citeauthoryear{MacCallum and Skea}{1994}]{MacSkea94}
M.A.H. MacCallum and J.E.F. Skea (1994) SHEEP:A computer algebra system for general relativity, in {\it Algebraic Computing in General Relativity} M. J. Rebou\c cas and W.L. Roque, editors, Clarendon Press, Oxford
\bibitem[\protect\citeauthoryear{\AA man {\it et al.}} {1997}]{Aman97}
J.E. \AA man, J.B. Fonseca-Neto, M.A.H. MacCallum and M.J. Rebou\c cas (1997) Computer-aided Classification of Riemann-Cartan Space-times of G\" odel Type
\bibitem[\protect\citeauthoryear{Pollney {\it et al.}} {1997}]{Pollney97}
D. Pollney, R. d'Inverno and J.E.F. Skea (1997) Programs for Invariant Classification in Maple
\bibitem[\protect\citeauthoryear{{\AA}man and Karlhede} {1981}]{Karlhede81}
J.E. {\AA}man and A. Karlhede (1981) An Algorithmic Classification of Geometries in General Relativity in {\it  Proceedings of the
1981 ACM Symposium on Symbolic and Algebraic Computation (Symsac '81)}
\bibitem[\protect\citeauthoryear{d'Inverno {\it et al.}} {1997}]{d'Inverno97}
J.E.F. Skea, D. Pollney and R. d'Inverno (1997) An On-line Database of Exact Solutions and Invariant Classifications
\bibitem[\protect\citeauthoryear{Kramer {\it et al.}} {1980}]{Kramer80}
D. Kramer, H. Stephani, E. Herlt, M.A.H. MacCallum and E. Schmutzer (1980) Exact Solutions of Einstein's Field Equations, CUP, Cambridge
\bibitem[\protect\citeauthoryear{O'Connor}{1997}]{O'Connor97}
J.E.R. O'Connor (1997) Finding Collineations of Kimura Metrics
\bibitem[\protect\citeauthoryear{H\"oglund}{1997}]{Hoglund97}
A. H\"oglund (1997) Tensign, the philosophy behind a new user interface
\bibitem[\protect\citeauthoryear{Edgar {\it et al.}}{1997}]{Edgar97}
S.B. Edgar and A. H\"oglund (1997), {\it Proc. R. Soc. Lond. A}, {\bf 453}, {pp.~835}
\bibitem[\protect\citeauthoryear{Andersson {\it et al.}}{1997}]{Andersson97}
 F. Andersson, S. B. Edgar and A. H\"oglund (1997) The Lanczos potential for the Weyl curvature tensor exists in four, and only in four dimensions. GR15 Abstracts: Workshop A3
\bibitem[\protect\citeauthoryear{Musgrave}{1997}]{Musgrave97}
P. Musgrave and K. Lake (1997) GRLite: GRTensor for the Web \footnote{Another Java-based virtual algebraic calculator has been developed recently by John Mourra. Follow www.astro.queensu.ca/\char'176lake/vac.html for more information.} 
\end{thebibliography}
\end{document}